\documentclass[12pt,a4paper]{article}
\usepackage{graphics,amsfonts}

\newcommand{\be}{\begin{equation}}
\newcommand{\ee}{\end{equation}}
\newcommand{\eel}[1]{\label{#1}\end{equation}}
\newcommand{\bea}{\begin{eqnarray}}
\newcommand{\eea}{\end{eqnarray}}
\newcommand{\eeal}[1]{\label{#1}\end{eqnarray}}
\newcommand{\baq}{\begin{equation}\begin{array}{rcl}}
\newcommand{\eaq}{\end{array}\end{equation}}
\newcommand{\eaql}[1]{\end{array}\label{#1}\end{equation}}
\newcommand{\beac}{\begin{equation}\begin{array}{rcl}}
\newcommand{\eeacn}[1]{\end{array}\label{#1}\end{equation}}
\newcommand{\ba}{\begin{array}}
\newcommand{\ea}{\end{array}}
\newcommand{\non}{\nonumber \\}
\newcommand{\equ}[1]{(\ref{#1})}

\newcommand{\beq}{\begin{eqnarray}}
\newcommand{\eeq}{\end{eqnarray}}
\newcommand{\nn}{\nonumber}

\newcommand{\adss}{$AdS_5\times S^5$ }
\newcommand{\adssns}{$AdS_5\times S^5$}
\newcommand{\pa}{\partial}
\newcommand{\pu}{\partial_u}
\newcommand{\pat}{\partial_t}
\newcommand{\px}{\partial_x}
\newcommand{\preprint}[1]{\begin{table}[t]  
           \begin{flushright}               
           \begin{large}{#1}\end{large}     
           \end{flushright}                 
           \end{table}}                     

\def\half{{1\over 2}}

\def\cd{{\cal{D}}}

\def\czt{{{\cal{Z}}_{(2)}}}
\input{psfig}
\begin{document}

\begin{titlepage}

\preprint{hep-th/9911123\\TAUP-2602-99}

\vspace{2cm}

\begin{center}
{\bf\LARGE  Quantum fluctuations of Wilson loops from string
models }\footnote{
Work supported in part by the US-Israel Binational Science
 Foundation,
by GIF - the German-Israeli Foundation for Scientific Research,
and by the Israel Science Foundation.
}\\
\vspace{1.5cm}
{\bf Y. Kinar  $\;\;$ E. Schreiber $\;\;$ J. Sonnenschein}\\
\vspace{.4cm}
{\em Raymond and Beverly Sackler Faculty of Exact Sciences\\
 School of Physics and Astronomy\\
 Tel Aviv University, Ramat Aviv, 69978, Israel\\
 e-mail: \{yaronki,schreib,cobi\}\verb+@+post.tau.ac.il}

\vspace{.5cm}
{\bf N. Weiss}\\
\vspace{.4cm}
{\em Department of Physics\\
University of British Columbia,\\
Vancouver, B.C., V6T2A6, Canada\\
e-mail: weiss@physics.ubc.ca}
\end{center}
\vskip 0.61 cm
\begin{abstract}

We discuss the  impact of  quadratic  quantum
fluctuations on the
 Wilson loop extracted from
classical string theory.
 We show that  a large class of models,  which includes the near horizon
limit of
 $D_p$ branes with 16 supersymmetries,
admits
a L\"{u}scher type correction to the classical potential.
 We confirm that the quantum determinant  associated with
a BPS
configuration of a single quark in the \adss model
is free from divergences. We find  that for  the Wilson loop in that model,
unlike the situation in flat space-time,
 the fermionic determinant does not
cancel the bosonic one. For string models
 that correspond to gauge
theories in the  confining phase,  we show that
 the correction to the potential is
 of a L\"{u}scher type and is attractive.

\end{abstract}

\end{titlepage}

\section{Introduction}

The idea of describing the Wilson loop of QCD in terms of a string partition
function dates back to the early Eighties.
In a landmark paper in this direction \cite{LUS} it was  found that
the potential  of quark anti-quark separated at a distance $L$ acquires
 a  $-{c\over L}$ correction term ($c$ is a positive universal constant
independent of the coupling)  due to
 quantum fluctuations of a Nambu--Goto (NG) like action.
This term is commonly called the L\"{u}shcer term.
 An exact expression of  the partition function of the NG action
was  derived in the large $d$ limit \cite{Alvarez}, where $d$ is the
space-time dimension, and the 1-loop and $1/d$ expansions were
considered for strings and $p$-branes in various space-time topologies
\cite{OdWi,ByOd}.
The large $d$ result,  when translated to the quark  anti-quark potential,
takes the form
$E(L)\sim {L\over \alpha'}\sqrt{1-{2c\over L^2/\alpha'}}$.
Thus, by  expanding it  for small  ${2c \over L^2/\alpha'}$,
one finds  the linear confinement potential  as well as the
L\"{u}scher term.  It was further shown that in this approximation
the semiclassical potential associated with  Polyakov's action and NG
action are identical \cite{FraTse}.
It was later realized that  in fact this expression
is identical to the energy of the tachyonic  mode of the bosonic string
in flat spacetime with Dirichlet boundary conditions at $\pm L/2$
\cite{Arvis}.

Recently there has been a Renaissance of the
 idea  of a stringy description of the Wilson loop in the framework
of Maldacena's correspondence between large $N$ gauge theories and
string theory \cite{Mal1}.
Technically, the main difference between the "old" calculations and the
modern ones is the fact that the spacetime background is no longer flat
but rather an \adss or certain generalizations of it.
Conceptually, the modern  gauge/string  duality gave the stringy description
a more solid basis.
The first "modern" computation \cite{Mal2, ReYe} was for the \adss metric
which corresponds to the ${\cal N}=4$ supersymmetric theory.
 To make contact with non-supersymmetric gauge dynamics one makes use of
Witten's idea \cite{Witads2} of putting the Euclidean time direction on a
circle
with anti-periodic boundary conditions.
This recipe was utilized to determine the behaviour of the potential for
the  ${\cal N}=4$ theory at finite temperature \cite{RTY,BISY1}
 as well as 3d pure YM theory \cite{BISY2}
which is the limit of the former at infinite temperature.
Later, a similar procedure was invoked to compute Wilson loops of 4d YM
theory,
 `t Hooft loops \cite{BISY2, GroOog} and  the quark anti-quark potential
in MQCD and in Polyakov's type 0 model \cite{KSS1,Polyakov,KSS2}.
A unified scheme for all these models and variety of others was analyzed
in \cite{KSS2}. A theorem  that determines the leading and next to
leading behaviour of the classical potential associated with this unified
setup was proven and applied to several models. In particular a corollary
of this theorem states the sufficient conditions  for the potential to have
a confining nature.

The issue of the quantum fluctuations and the detection of a L\"{u}scher
term
was raised again in the modern framework in \cite{GrOl1}. It was noticed
there
that a more accurate evaluation  of the classical result \cite{KSS2}
did not have
 the form of a L\"{u}scher term. This is, of course, what one should have
anticipated, since after all  the  origin of the L\"{u}scher term \cite{LUS}
is the quantum fluctuations of the NG like string.
The  determinant associated  with
the bosonic quantum fluctuations of the pure YM setup was  addressed
in \cite{GrOl2}. It was shown
there that the system is approximately described by six operators
that correspond to massless bosons in flat spacetime and two additional
massive modes. The fermionic determinant was not computed in this paper.
However, the authors raised the possibility that the
latter will be of the form of massless   fermions   and hence there might
be a violation of the concavity behaviour of gauge potentials \cite{Bachas,
DorPer}. One of the results of our work is that in  fact the
fermionic operators are massive ones and thus the bosonic determinant
dominates and there is an attractive interaction after all.
The impact of the quantum fluctuations  for the case of the \adss case
was  discussed in \cite{FGT, Naik}. Using the GS action
\cite{MetTse,Pesando,KalRah,KalTse}
with a particular $\kappa$ symmetry fixing,  it was observed that the
corresponding quantum Wilson loop suffers from UV logarithmic divergences.
It was argued that by renormalizing the mass of the quarks one can remove
the divergence.

The computations of Wilson loops in 3d and 4d pure YM theory  can be
confronted with the results found in lattice calculations. In  particular,
the main question is whether  the correction
to the linear potential   in the form of a L\"{u}scher term can be detected
in  lattice simulations.
According to \cite{Teper} there is some numerical evidence for a L\"{u}scher
term associated with a bosonic string, however the results are not
precise enough to be convincing.
Obviously, our ultimate dream is  compatibility with heavy meson
phenomenology.

In this paper our main goal has been  the quantum corrections of  the
quark anti-quark potential in the class of
non-supersymmetric confining theories. On the route to this target
we had to overcome
several related obstacles.  An important technical problem  is the issue of
gauge fixing. Whereas for flat space-time backgrounds   there are
several known fixing procedure which are under full control, it turns out
that for the non flat background   the situation is more subtle.
 Already in fixing the world sheet diffeomorphism  we were facing
gauge choices that looked innocent but later were found to be problematic.
The
situation with the fixing of the $\kappa$--symmetry is even
trickier.  We wrote the form of the bosonic action truncated to quadratic
fluctuations  for the unifying scheme of \cite{KSS2}.
As warm up exercises we considered the fluctuations of a string in
flat space-time and  the fluctuations of a BPS quark of the \adss model.
Whereas in the former case it is quite straightforward to realize that the
bosonic part yields the L\"{u}scher term and the fermionic part exactly
cancels
it, in the later case the picture is more involved.
Eventually, the basic expectation  that there are no divergent
corrections to the energy  was verified. However,  this result emerged only
in a particular class of $\kappa$--symmetry  fixing schemes.
Without performing an explicit computation,
using a scaling argument  we were able to
write down the general $L$ dependence for a large class of models. It turns
out that the  set  of models based on $D_p$ branes with 16 supersymmetries
have a L\"ushcer type behaviour. Note that unlike the $p=3$ case, for
$p\neq 3$ there is no reason from dimensional grounds for the potential
to be of the form $1/L$, and indeed the classical result is not of this
form.

As for the case of the Wilson loop of the \adss model we found
that there is only partial cancellation between the bosonic and fermionic
determinant. The net free energy  is of a L\"{u}scher form, but
we were unable to determine neither its coefficient nor even its sign.

In spite of the fact that there is no GS action which corresponds
 to the non flat background associated with confining gauge theories,
we argue that there is a net attractive L\"{u}scher type correction
to the linear potential in the pure YM case.

The outline of the paper is as follows. We start  with a brief description
 of the classical setup. For that purpose we use the Nambu Goto action
associated with a space-time metric which is diagonal and depends on only
one
coordinate. The \adss is a special case of this configuration and so are
backgrounds that correspond to confining gauge dynamics. In section 3 we
introduce   quantum  fluctuations to the string coordinates.
We expand the action to quadratic order in the fluctuations and
 write down the general form of the operators whose  (log) determinant
determines  the free energy of the system.
When translating to the gauge language the free energy has the
interpretation
of the quantum correction of  the classical quark anti-quark potential.
We discuss  three possible gauge fixing schemes and argue that only one of
them, the "normal coordinate fixing" is  a "safe" gauge.
 As a warm--up exercise of computing and renormalizing the determinant, we
derive  in section 4 the L\"{u}scher term in the flat space-time background.
Section 5 is devoted to  an analysis of the
dependence of the free energy on the separation distance $L$
based on a  general scaling argument. An application of this result to the
case of $D_p$ branes with 16 supersymmetries \cite{IMSY} shows that
the quantum correction for this configuration must also be of
L\"{u}scher type.
We then use, in section 6, the general expressions of the operators for
the \adss case. We rederive these  results in  the framework of
Polyakov's action
in section 7. The next task of incorporating the fermionic fluctuations
is discussed in section 8, again first in the simplest case of flat
space-time.
We then proceed to the case of a single BPS quark in the \adss background.
We discuss  the fixing of the $\kappa$--symmetry and show that there
are certain subtle issues involved.  Eventually, using a theorem about
Laplace type operators, we show In section 9 that the BPS free energy
is free from
divergences due to cancellations between the bosons and fermions.
The quantum fluctuations of the Wilson loop in the \adss background or
its generalization to the cases of $D_p$ with 16 supersymmetries  are
analyzed in section 10. We argue that there is left over uncanceled
L\"ushcer
term but we are unable to determine neither its coefficient nor its sign.
Sections 11--13 are devoted to a similar analysis in  the setup
corresponding to a confining behaviour.  In section 11 We show that
the general case
of this type behaves in a similar manner to the "pure gauge configuration",
namely, that the Wilson line   is well approximated  by a straight string
parallel to the horizon and close to it.
For this type of backgrounds, in spite of the fact that we do not have
a detailed Green Schwarz action  we were able to show that the residual
interaction is of an attractive nature. The Bosonic operators are
considered in section 12, while the fermionic ones are considered in
section 13.
We end this paper with a summary of our results and a list of open
questions.

\section { The classical setup }
\label{sec:setup}

Let us first, before introducing quantum fluctuations, summarize the
results derived from classical string theory. Having in mind
 space-time backgrounds associated with the gravity/gauge dualities
 and MQCD, we  restrict ourselves to diagonal background metrics  with
  components  that depend only on  one coordinate. The metric
  $G_{\mu\nu}$ for the coordinates $\{t,x,y_i,u, \zeta_I \}$
where $i=1,...,p-1$ and $I=p+2,...,9$ is given by
\be
       G_{\mu\nu} = \alpha' \times
diagonal\{G_{tt}(u),G_{xx}(u),G_{y_iy_i}(u),
G_{uu}(u), G_{\zeta_I \zeta_I }(u)\}
\ee

The classical (and, as we shall see, the one-loop) bosonic string has
two equivalent descriptions in terms of the Nambu--Goto action  and
 the Polyakov action. Throughout this work we incorporate quantum
 fluctuations mainly in  the former picture. Comparison  with the
 analysis in the Polyakov formulation is presented in section
 \ref{sec:polyakov}.

The Nambu--Goto action of a string propagating in space--time with a
 general background metric takes the form

\be
\label{NGAct}
      S= {1\over 2\pi\alpha'}\int d\sigma d\tau \sqrt{\det \:
h_{\alpha\beta}}
\ee
where

\be
      h_{\alpha\beta} = \partial_\alpha X^\mu \partial_\beta X^\nu
G_{\mu\nu}(X)
\ee

The classical string configurations associated with Wilson loops
are characterized by boundary conditions  which take the form of a loop
on a boundary plane. The plane  is taken  at some large but finite value
$u=u_s$ which
 later is  taken  to infinity. The form of the loop considered in the
present work is that of a very long strip  along  a spatial direction
($-L/2<x<L/2$) and a temporal length $T$ with $L<<T\rightarrow\infty$.
In certain models we will take the ``temporal" direction to be along
another spatial direction.

Since we assume invariance under time translation, we consider
classical static string solutions. These are  spanned in the $x,
u$ plane by the function $u(x)$ and we set $y_i = 0, \; \zeta_I = 0$.

The NG action is invariant under world-sheet coordinate transformation.
It is convenient to use the static gauge in which we set the  world-sheet
time to be identical to that of the target space,  $\tau = t$. Various
different fixings of $\sigma$ will be discussed in the next section
when quantum fluctuations are incorporated. For the classical
configuration we take  $\sigma = 2\pi x$. Defining the
 two functions
\beq
       f^2(u) & \equiv & G_{tt}(u) G_{xx}(u) \\
       g^2(u) & \equiv & G_{tt}(u) G_{uu}(u) \nn
\eeq

the action  becomes

\be\label{NGact}
      S = T \int {\cal L} \, dx = T \int dx  \sqrt{f^2(u) + g^2(u) \:
(\partial_x
u)^2}
\ee

Using the fact that the Lagrangian does not depend
explicitly on $x$ we write the equation of motion in terms
of the conserved  generator of translations along $x$ so that
\be
\label{dxucl}
      \partial_x u_{cl} = \pm {f(u_{cl}) \over g(u_{cl})} \cdot
{\sqrt{f^2(u_{cl}) - f^2(u_0)} \over f(u_0)}
\ee
where $u_0$ is the minimal value of $u$ reached by the string.

The  classical profile of the  Wilson loop is a solution of \equ{dxucl}
which satisfies the boundary conditions stated above.
Since, in the language of the corresponding gauge model, the  quark
and anti-quark   are set at the coordinates $x=\pm L/2$, the relation
between $L$ and $u_0$ is given by

\be
L = 2 \int_{u_0}^{u_s} \frac{g(u)}{f(u)}
\frac{f(u_0)}{\sqrt{f^2(u)-f^2(u_0)}} du
\ee
and  the action is given by
\be
\label{Eofu0}
S = 2 T\int_{u_0}^{u_s} \frac{g(u)}{f(u)}
\frac{f^2(u)}{\sqrt{f^2(u)-f^2(u_0)}} du
\ee

The basic conjecture of the string/gauge duality is that the
``natural candidate" for the expectation value $\langle W \rangle$ of
the Wilson loop is proportional to the partition function  of the
corresponding string action
\be
\label{wdef}
        \langle W \rangle \propto {\bf Z}=\int \cd X^\mu \exp{(-S)}
\ee
where the integral is over all surfaces whose boundary is the given loop.
Moreover,  $\langle W \rangle$ is related to the quark anti-quark potential
energy $E(L)$  via
\be
        \langle W \rangle \rightarrow\exp\left({-TE(L)}\right)
\ee
so that in fact \equ{Eofu0} determines the potential in the classical
limit in which the configuration of least action dominates.
However, it is easy to see that the expression  \equ{Eofu0} is
linearly divergent as $u_s\rightarrow \infty$ and hence has
 to be renormalized. The mass of the quarks which translate
in the string language into a straight line between $u=0$  (or $u=u_h$
 in case there is an horizon at $u_h$ \cite{BISY1}), is  given by
$m_q=\int_{0}^{u_s}g(u) du$. It is thus ``physically natural" to
determine the quark anti-quark potential as $E= {S\over T}- 2 m_q$.
For the special case of the  \adss model this definition of the energy
matches  the Legendre transform of the action suggested in \cite{DGO}.

By implementing this  expression for various different background metrics,
Wilson loops corresponding to certain gauge systems were computed
\cite{KSS2}.
Moreover, a theorem was stated in \cite{KSS2}   determining the
dependence of $E$ as a function of $L$ for the general setup of \equ{NGact}.

\section {Quadratic  fluctuations and gauge fixings in the NG action}
In order to account for the contributions to the Wilson loop from
quantum fluctuations we expand  the coordinates  around their classical
values
\be
x^\mu(\sigma,\tau) = x^\mu_{cl}(\sigma,\tau) + \xi^\mu(\sigma,\tau)
\ee
that is,
\be
\label{Xflucts}
\{t=t_{cl} + \xi_t \;;\; x = x_{cl} + \xi_x \;;\; u = u_{cl} + \xi_u \;;\;
y_i = \xi_{y_i} \;;\; \zeta_I=\xi_{\zeta_I} \}
\ee
The next step is to expand the Nambu--Goto action  up to terms quadratic
in the fluctuations  so that $S=S_{cl}+S_{(2)}$, and compute the following
path integral
\be
\label{pathint}
   \czt=\int \cd \xi_t  \cd \xi_x \cd \xi_u  \prod \cd \xi_{y_i}
     \prod \cd \xi_{\zeta_I}        \exp \left\{ -S_{(2)}\right\}
\ee
where we write
\be
\label{S2}
S_{(2)} = \sum_a \int d\sigma d\tau \; \xi_a^\dagger {\cal O}_a \xi_a
\ee
with $\xi_a$ the various fields.
\footnote{Our treatment differs from that of
\cite{FGT} in that the authors of that article explicitly include in
the integrals of \equ{S2} the measure $ (\det h_{\alpha \beta}^{(0)})^{1/2}$
corresponding to the classical solution, and change the operators
accordingly.}

The Gaussian integration over the fluctuations yields ( after gauge
fixing) a product of determinants of  second order differential
operators $\prod_a \det{\cal O}_a$ so that the corresponding free
energy is given by
\be
F_B = -\log \czt= -\sum_a\half \log \det {\cal O}_a
\ee

As was mentioned above, the NG action is invariant under world--sheet
reparameterizations, and in order to compute explicitly \equ{pathint}
one has to introduce a gauge choice.
In fact, without gauge fixing the operators are degenerate due to the
reparameterization invariance \cite{KalTse}. Obviously, the set of
 operators ${\cal O}_a$ depends on the gauge. In the following
 subsections we write down  the form of the operators in three
 different gauges. We set  in all the three gauges $\tau=t$ so there
are no fluctuations in the time direction. The gauges differ in the
way we fix $\sigma$ and the choice of the fluctuating coordinates.
 In the first gauge  we fix the $u$ coordinate and compute the
fluctuations of $x$, $y_i$ and $\zeta_I$ while in the  second one
  $x$ is fixed and $u$, $y_i$ and $\zeta_I$ fluctuate.
  In the third gauge   the fluctuations are taken to be  along the normal
coordinate to $u_{cl}$ (in the $ux$ plane)  and along $y_i$ and $\zeta_I$.

\subsection{ The fixed $u$ gauge}

Imposing the assignment  $\sigma = u$ (running from $u_0$ to $\infty$)
one finds that  to the second order in the fluctuations
$\xi_x$ and $\xi_{y_i}$, the expression $h = \det \: h_{\alpha\beta}$
takes the form
\beq
h_{(2)} &= & g^2(u) \left(1 + {G_{xx}(u) \over G_{tt}(u)} (\pat \xi_x)^2 +
           \sum_i {G_{y_i y_i}(u) \over G_{tt}(u)} (\pat \xi_{y_i})^2
\right) \non
          &+& f^2(u) \left((\pu x_{cl})^2 + 2(\pu x_{cl})(\pu \xi_x) +
            (\pu \xi_x)^2 \right) \\
&+&
f^2(u)\sum_i \left( {G_{y_i y_i}(u) \over G_{xx}(u)} (\pu \xi_{y_i})^2 +
              {G_{y_i y_i}(u) \over G_{tt}(u)} (\pu x_{cl})^2 (\pat
\xi_{y_i})^2 \right)
 \nn
\eeq
where $x_{cl}(u)$ is defined by $u_{cl}(x_{cl}(u)) = u$.

Simplifying the last two terms (Using \equ{dxucl}) and integrating by parts,
we find
that the set of  operators  ${\cal O}_a$ includes two types of
quadratic operators:

\beq
       {\cal O}_x &=& \half \left[\pu
         \left(\frac{(f^2(u)-f^2(u_0))^{3/2}}{f(u) g(u)} \pu\right) +
         \sqrt{f^2(u)-f^2(u_0)} {g(u) \over f(u)} {G_{xx}(u) \over
           G_{tt}(u)} \pat^2 \right] \non
       \label{Oy(u)}
       {\cal O}_{y_i} &=& \half \left[\pu \left({G_{y_i y_i}(u) \over
             G_{xx}(u)} \sqrt{f^2(u)-f^2(u_0)}\,{f(u) \over g(u)}
           \pu\right) \right. \\
       & & \hspace{5cm} +  \left. {G_{y_iy_i}(u) \over G_{tt}(u)}
{{g(u)f(u)}
           \over \sqrt{f^2(u)-f^2(u_0)}} \pat^2 \right] \nn
\eeq
and similarly for ${\cal O}_{\zeta_I}$ with $G_{\zeta_I\zeta_I}(u)$
replacing $G_{y_iy_i}(u)$.
The boundary conditions we impose on the eigenfunctions are
$\xi(u,0)=\xi(u,T)=0$ , $\xi(\infty,t)=0$ and $\xi(u_0,t)=0$ or
$\xi'(u_0,t)=0$
(The conditions at $u_0$  imply that the function $\xi$ should be
symmetric or antisymmetric around $x=0$).

For a $p+1$ dimensional space-time $(t,x,y_i)$, the bosonic free
energy is given by:
\be
\label{FrEn}
       F_B = - \half \log {\det {\cal O}_{x}}
- {{p-1} \over 2} \log {\det {\cal O}_{y}}
- {(8-p)\over 2} \log {\det {\cal O}_{\zeta}}
\ee

We can change variables (i.e. change world-sheet coordinates) to
$\sigma=x_{cl}$ and obtain the quadratic operators:
\beq
\label {QuadOp}
      \hat{\cal O}_x &=& {f(u_0) \over 2} \left[\px\left((1-{f^2(u_0) \over
f^2(u_{cl})})
\px\right) + {G_{xx}(u_{cl}) \over G_{tt}(u_{cl})} ({f^2(u_{cl}) \over
f^2(u_0)}-1)\pat^2
\right] \non
      \hat{\cal O}_{y_i} &=& {f(u_0) \over 2} \left[\px\left({G_{y_i
              y_i}(u_{cl}) \over G_{xx}(u_{cl})} \px\right)
+ {G_{y_i y_i}(u_{cl}) \over G_{tt}(u_{cl})} {f^2(u_{cl}) \over f^2(u_0)}
\pat^2 \right]
\eeq
and similarly for $\hat{\cal O}_{\zeta_I}$
where the boundary conditions are $\hat\xi(-L/2,t) =
\hat\xi(L/2,t)=0$. The free energy is still given by \equ{FrEn}.

\subsection{ The normal coordinate gauge}
\begin{figure}[h!]
\begin{center}
\resizebox{0.6\textwidth}{!}{\includegraphics{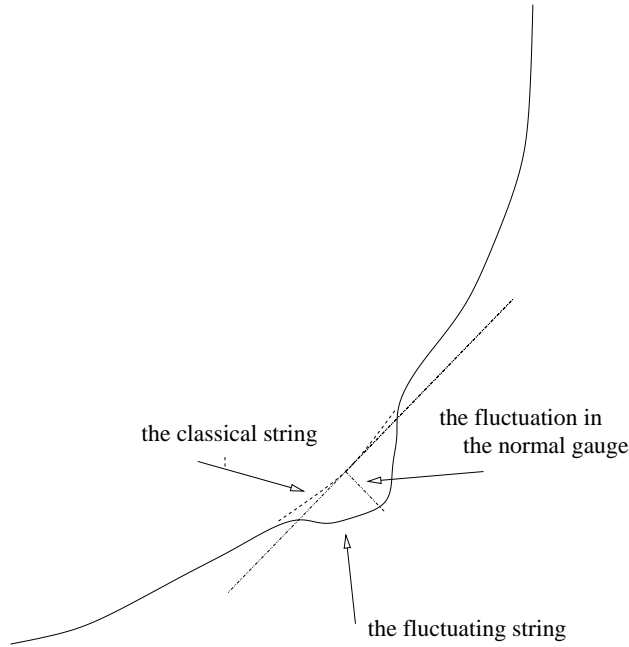}}
\end{center}
\caption{The fluctuations in the normal gauge}
\label{fig:normal}
\end{figure}

In this gauge we take the
fluctuations to be everywhere normal to the classical curve (see figure
\ref{fig:normal}). With this
choice, both $u_{cl}$ and $x_{cl}$ are changed by  the fluctuation to the
final values $u$ and $x$.
We further choose the gauge  $\tau = t$ (as before), and
$\sigma = u_{cl}$. The components of the tangent vector to the
classical curve obey $t_u/t_x = \partial_x u_{cl}$ which is given by
\equ{dxucl}, and therefore the components of the unit normal vector are the
solutions of
\begin{eqnarray}
n_x(u_{cl})\, t_x(u_{cl})\; G_{x x}(u_{cl})  +
n_u(u_{cl})\, t_u(u_{cl})\; G_{u u}(u_{cl})    & = & 0 \non
n_x(u_{cl})^2 \; G_{x x}(u_{cl}) +
n_u(u_{cl})^2 \; G_{u u}(u_{cl})   & = & 1
\end{eqnarray}

We denote the magnitude of the fluctuation along the normal direction
by the dimensionless field $\xi_n(u_{cl},t)$. The coordinate vector
in the $(x,u)$ plane is thus
\be
      x= x_{cl}(u_{cl}) + n_x(u_{cl}) \, \xi_n(u_{cl},t) \;,\;
                          u=   u_{cl} + n_u(u_{cl}) \, \xi_n(u_{cl},t)
\ee
and both $u_{cl}$ and $x_{cl}$ change after the fluctuation to the
final values $u$ and $x$.

Unlike the fixed $u$ gauge, the fluctuations result also in a change
in the metric. In our gauge the off-diagonal elements are zero and
the diagonal elements, to second order, are given by:
\beq
G_{\mu \mu}(X) &=& G_{\mu \mu}(u_{cl}) +
                 \partial_u G_{\mu \mu}(u_{cl})
                   \cdot n_u(u_{cl}) \, \xi_n(u_{cl},t) \non
               &+&  \frac{1}{2} \partial_u^2 G_{\mu \mu}(u_{cl})
                   \cdot \left(n_u(u_{cl}) \, \xi_n(u_{cl},t)\right)^2
\eeq
(with no summation over indices).
As the action is an integral of ${\cal L}$ essentially from
$u=\infty$ to $u=u_0$ and back again to $u=\infty$, and we have the
boundary conditions $\xi_n(\infty,t) = 0$, the part of ${\cal L}$
linear in $\xi_n$ does not have to vanish, but rather should be
a total derivative of the form
$\partial_u (F_{(1)}\, \xi_n(u_{cl},t))$, in order to ensure that the
classical solution is an extremum. We find this to be indeed the case.
Inspection of $h_{\alpha \beta}$ readily reveals that
there are in $h$ no quadratic terms of the types $\partial_t \xi_n \cdot
\xi_n$
or $\partial_t \xi_n \cdot \partial_u \xi_n$ (and also no linear term of
the type $\partial_t \xi_n$). By adding a total derivative of the type
$\partial_u (F_{(2)}\, \xi_n(u_{cl},t)^2)$ we can also get rid of the
$\partial_u \xi_n \cdot \xi_n$ term. Finally, we get the quadratic
expression
\be
\label{Llongnormal}
{\cal L}_{(2)} = {\cal L}_{cl} +  C_{\xi \xi}(u_{cl}) \xi_n^2 +
C_{t t}(u_{cl}) (\partial_t \xi_n)^2 + C_{u u}(u_{cl}) (\partial_u \xi_n)^2
\ee
Thus the  operator ${\cal O}_n$ takes the form
\be
\label{GenOpr}
{\cal O}_n= -\partial_u( C_{u u}(u_{cl})\partial_u) -
                         C_{t t}(u_{cl}) \partial_t^2
                      +  C_{\xi \xi}(u_{cl})
\ee
Since the explicit expressions of the $C$'s  are cumbersome  for the general
case
we do not find it useful to write them down.
In sections \ref{sec:bosonadss}, \ref{sec:Bosonpym}
 we  derive them  for certain   special cases.

\subsection {The fixed $x$ gauge}

Here we impose $\sigma = x$  and fix the gauge by $\xi_x=0$.
Expanding now the square root of $h$ to quadratic order yields
 yet another
operator for the longitudinal fluctuations. We write it this time
in  the form of the action.
$$
        S_{(2)}={f(u_0)\over 2}\int d\sigma d\tau \left\{\left[
        {{g(u)^2 f(u_0)^2}\over{f(u)^4}}\left(\partial_\sigma\xi_u\right)^2
        +{G_{uu}({u})\over      G_{tt}({u})}
        \left(\partial_\tau\xi_u\right)^2
        + \mu^2(\sigma)\xi_u^2 \right]
        \right.
$$
\be \label{xeqsigquad}
        \left. +\left[{G_{y_i y_i}({u})
        \over G_{xx}({u})}
        \left(\partial_\sigma\xi_i\right)^2
        +{{G_{y_i y_i}({u})G_{xx}({u})}\over f(u_0)^2}
        \left(\partial_\tau\xi_i\right)^2
        \right]
        \right\}
\ee
The "mass term" $\mu^2(\sigma)$ will be written explicitly
for  specific metrics.
The fact that there  is  a  "mass"
term
 present in this gauge,
 and in the normal coordinate gauge,  and there is no such term in the
fixed $u$ gauge
 is not surprising since such a term necessarily
arises when $G_{\mu\nu}(u)$ is expanded about ${u_{cl}}$. In the gauge
$\sigma=u$ there is no fluctuation in $u$ so no such term arises. The
significance of this term will be discussed further when we study special
cases such as AdS and pure Yang-Mills.



\subsection{ Comparing the different gauges}
One expects that physical, gauge invariant quantities should be identical
when using different gauges.  This general belief  is justified  provided
that
one considers legitimate gauges.
The differential operators derived in the three different gauges
are manifestly not identical. It may happen that their determinant
is nevertheless identical.
Formally,  one can pass from one gauge to another. For instance changing
the integration variable from $\sigma = x$ to $\sigma = u$  together with
the rescaling of  $\xi_u$
\be
        \label{scalinga}
        \xi_u = \xi_x \cdot \partial_x {u}(x)
\ee
 maps  between the fixed $x$ gauge to the fixed $u$ one.
However, this rescaling itself may be singular as indeed occurs at $u_0$.

\begin{figure}[h!]
\begin{center}
\resizebox{0.6\textwidth}{!}{\includegraphics{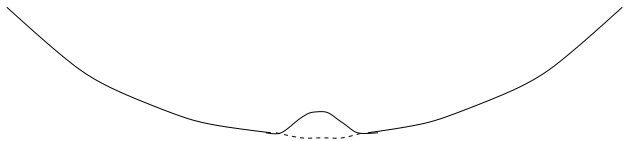}}
\end{center}
\caption{A problematic fluctuation in the fixed $u$ gauge}
\label{fig:problem}
\end{figure}

The computation of the operator ${\cal O}_x$  in the fixed $u$ gauge
corresponds to the
assumption that the longitudinal fluctuation of the string is in
the $x$ direction ("gauge fixing" the coordinates $u,t$ as
$\sigma,\tau$ respectively). This assumption leads to two problems
in the physical interpretation. First, near $u=u_0$, even a small
fluctuation can lead to multiple valuedness of $x$ as a function of
$u$. Second, the extremal point can itself fluctuate in the $u$
direction, and no longer correspond to $u=u_0$ (see figure
\ref{fig:problem}). The problem is that at $u=u_0$ the $x$ direction is
tangential to the classical curve, and that choice of the direction of
the field of fluctuations becomes singular. As the metrics depend only
on $u$, this choice of fluctuation involves no change in metric. It is
easy to understand, therefore,  why the operator ${\cal O}_x$
involves only derivatives of the fluctuating field.

By choosing $\sigma$ to be
$x$, and the direction of fluctuations to be along $u$, the limit
$u = \infty$ becomes singular in the aforementioned sense. However,
this singularity exists also in the fluctuations of the bare quarks,
and after the subtraction of the quark masses from \equ{Eofu0} it
might cancel.

The safest choice, however, seems to be taking the
fluctuations to be anywhere normal to the classical curve, as
previously explained.
\footnote{In \cite{FGT} the computations were performed in the "Riemann normal
coordinate" approach. This method is similar to the normal coordinate
we are using.}

\section{ Bosonic string in flat space-time}
\label{sec:luscher}
We can now easily derive the free energy of a bosonic string
in flat space-time (L\"{u}scher term).
The metric is given by $G_{tt}=-1\; ; \; G_{xx} = G_{y_iy_i}=1$,
where $i=1,\ldots,D-2$.
Note that  $u$ is now
one of the $y_i$ coordinates.
The classical solution  in this case is just a straight line along the
$x$ direction from $-L/2$ to $L/2$.
Obviously, for this case the three different gauges are identical. We
use \equ{QuadOp} and regard only the transverse fluctuations
operator ($\hat{\cal O}_{y_i}$).
This operator is  the ordinary Minkowski Laplacian. Considering now
Euclidean space time and demanding the eigenfunctions to vanish on the
boundary, the
eigenvalues of the Laplacian  are (minus)
\be
       E_{n,m} = ({{n \pi \over L}})^2 + ({{m \pi} \over T})^2
\ee
and the free energy is given by
\beq
       -\frac{2}{D-2} F_B & = & \log \det \Delta = \log \prod_{n,m} E_{n,m}
=
\sum_{n,m} \: \log (({n \pi \over L})^2 + ({m \pi \over T})^2) \\
         &\sim& \sum_n \: {T \over \pi} \int_0^\infty d\omega \log(1 + ({n
\pi \over \omega L})^2) + o(L) = \sum_n {T \over \pi} \cdot \pi \cdot {n \pi
\over L} + o(L) \non
         & = & T {\pi \over L} \sum_n n + o(L)
\eeq
where we assumed that $T$ is large.

Regulating this equation using the Riemann zeta
function and discarding (infinite) terms that do not depend on L, we find
\be
       F_B = (D-2){\pi \over 24} {T \over L}
\ee

This expression also gives a quantum correction to the linear quark
anti-quark potential in a bosonic QCD-string model in D-dimensional flat
space. Using the Wilson loop to calculate this potential we have
\be
      \Delta V(L) = - {1 \over T}  F_B = -(D-2) {\pi \over 24} \cdot {1
\over L}
\ee
which is called the L\"{u}scher term.

\section{A General scaling result}
\label{scaling}
The argument of the previous section can be generalized to operators
which do not necessarily correspond to the flat case and will lead
to a general
scaling result that will prove useful in our work.

Let us define the operators
\be
\label{OAB}
{\cal O}[A,B] = A^2 F_t(v) \partial_t^2 + B^2 {\cal O}_v
\ee
with $t$ the time, $v$ a coordinate whose range is independent of $L$.
${\cal O}_v$ is a general operator of $v$ alone, independent of $L$,
and $A,B$ are constants which may depend on $L$.
Then, $V[A,B]$, defined as the correction to the potential
arising from ${\cal O}[A,B]$, is proportional to $B/A$.
In particular, the potential is independent of any overall factor
multiplying the operator ${\cal O}$.

The idea of the proof is that by redefining $t$ we can
scale the operator ${\cal O}[A,B]$ to ${\cal O}[1,1]$, and the factors
by which the eigenvalues get multiplied in this operation become
irrelevant in the large $T$ limit. We proceed to the proof itself.

Separation of variables shows that the eigenfunctions of ${\cal
  O}[A,B]$ are $\phi(v,t) = e^{i \omega t} \phi(v)$.
We define the eigenvalues $\tilde{E}_n(\omega')$ by
\be
\left[ -F_t(v) \,\omega'^2 + {\cal O}_v \right]
\tilde{\phi}(v) = \tilde{E}_n(\omega') \, \tilde{\phi}(v)
\ee
and $E_n(\omega')$ by
\be
\left[ -A^2 F_t(v) \, \omega'^2 + B^2 {\cal O}_v \right]
\phi(v) = E_n(\omega') \, \phi(v)
\ee
or
\be
\left[ -F_t(v)(A \omega'/B)^2 + {\cal O}_v) \right]
\phi(v) = B^{-2} E_n(\omega') \, \phi(v)
\ee
so that
\be
E_n(\omega') = B^2 \tilde{E}_n(A \omega'/B)
\ee

The boundary conditions for the time coordinate lead to $\omega =
\frac{\pi m}{T}$ for integer $m$.
Hence we have for $V[A,B]$, (when $o(T)$ designates a term which does
not depend on $T$, or grows sublinearly with it),
\begin{eqnarray}
-T \cdot V[A,B] + o(T) & = & \frac{1}{2} \log \det {\cal O}[A,B] \non
  & = & \frac{1}{2} \sum_n \sum_m \log E_n(\frac{\pi m}{T}) \non
  & = & \frac{1}{2} \sum_n \sum_m
              \left(\log \tilde{E}_n(\frac{A}{B} \cdot \frac{\pi m}{T}) +
                   2 \log B \right) \non
  & = & \frac{1}{2} \sum_n \sum_m
              \left(\log \tilde{E}_n(\frac{\pi m}{B T / A})\; +\;
                   o(T) \right) \non
  & = &  -\frac{B T}{A} \cdot V[1,1] + o(T)
\end{eqnarray}
and so indeed $V[A,B] = (B/A) \cdot V[1,1]$.

We can make a simple consistency check of our result. Let us look at
the flat Laplacian $\Delta = \partial_{t}^2 + \partial_{x}^2$.
The range of $v = x/L$ is $0 \leq v \leq 1$ independent of $L$, and
$\Delta = \partial_{t}^2 + L^{-2} \partial_{v}^2$, so by the above result
we get $V(L) \propto L^{-1}$ as we have indeed seen in section
\ref{sec:luscher}.

\section{ Bosonic string in the \adss background}
\label{sec:bosonadss}
In \adss  background the metric can be written in the following form
\be
\label{adssmet}
        -G_{t t} = G_{x x} = G_{y y} = {u^2 \over R^2} ;
        ~~ G_{u u} = {R^2 \over u^2} ;
        ~~ G_{\zeta \zeta} = R^2
\ee
where $y$ is a transverse coordinate of the
$AdS_5$, while $\zeta$ is a transverse coordinate in $S^5$.
We therefore get
\beq
        f(u) &=& {u^2 \over R^2} \\
        g(u) &=& 1
\eeq
Note that as we deal with small fluctuations and big $R$,
the angular coordinates $\zeta$ of $S^5$ can be treated as if they were
not compactified.

Unlike the flat case, for the \adss metric the different gauge fixings yield
different operators.

In the fixed $u$ gauge one finds  using \equ{QuadOp} the following
quadratic operators
\be
      \hat{\cal O}_x = \half {u_0^2 \over R^2} \left[\px((1-{u_0^4
          \over u^4}) \px) -  ({u^4 \over u_0^4}-1)\pat^2 \right]
\ee
and
\beq
      \hat{\cal O}_{y} &=& \half {u_0^2 \over R^2} \left[\px^2 - {u^4
          \over u_0^4} \pat^2 \right] \non
\label{OpAds}
      \hat{\cal O}_{\zeta} &=& \half {u_0^2 \over R^2} \left[\px
        (\frac{R^4}{u^2} \px) - \frac{R^4 u^2}{u_0^4} \pat^2 \right]
\eeq

In the normal fluctuations gauge, the  fluctuations along  the directions
$y,\zeta$
which are perpendicular to the $(x,u)$ plane are the same as \equ{OpAds}.
However, for ${\cal O}_n$ we have
\be
\label{On}
{\cal O}'_n = \frac{2 u^8 - 5 u^4 u_0^4 + 3 u_0^8}{2 u^6 \sqrt{u^4 -
u_0^4}} -
              \frac{R^4}{2 \sqrt{u^4 - u_0^4}} \; \partial_t^2 -
              \partial_u \left(\frac{\sqrt{u^4 -
u_0^4}}{2}\,\partial_u\right)
\ee

In the dimensionless variable $v = u/u_0$, it is easy to see that this
operator is of the form \equ{OAB}, with $A^2 = R^4 u_0^{-2}, \; B^2 = 1$.
We conclude, according to the result of section \ref{scaling}, that the
correction
to the potential is proportional to $R^{-2} u_0$, that is \cite{Mal2},
inversely proportional to $L$. Moreover, there is no further
dependence on $R$ when we eliminate the variable $u_0$ in favour of
$L$. The classical and quadratic quantum quark anti--quark potential
add up to $V \sim -\sqrt{\lambda}/L + c'/L$ with
$\lambda \equiv g_{YM}^2 N \sim R^4$ the 't Hooft coupling constant, and
$c'$ a
universal constant, independent of $\lambda$. We see that the quantum
expansion in $\hbar$ is equivalent, for large $\lambda$, to an
expansion in $R^{-2}$ or equivalently in $\lambda^{-1/2}$.
The quantum expansion does not spoil, of course, the conformal nature
of the field theory. The other operators for the normal gauge can be
seen, by similar reasoning, to give rise also to a potential correction
of the same characteristics.

Note also that if we approximate  the classical curve to be flat over
most of its range and thus approximate $u=u_0$ throughout, the
mass term drops out and the operator simplifies to the flat Laplacian,
\be
\hat{\cal O}'_x \longrightarrow
        -\frac{1}{2}\,\partial_t^2 - \frac{1}{2}\,\partial_x^2
\ee
and for the longitudinal direction we get exactly the same L\"{u}scher
term as in the flat case. The same happens also to the transverse
directions. (Note that ${\cal O}_n^\prime$ cannot be approximated in this
way
since $u$ is constant and it is thus not integrated over ).
The shape of the string, however, does not change with $L$ but  only
scales. It is thus not a good approximation, in  our case,
to assume that the string is flat.
In section \ref{flat} we will give quite a
stringent criterion for flatness, and prove that strings in metrics
corresponding to confining field theories are indeed flat.

When we revert to a differential operator involving the coordinate $x$
instead of $u$, (with the integration of the Lagrangian in the $t,x$
coordinates), the operator \equ{On} is translated to
\be
\hat{\cal O}'_n = \frac{2 u^8 - 5 u^4 u_0^4 + 3 u_0^8}{2 R^2u^4 u_0^2} -
                  \frac{R^2 u^2}{2 u_0^2} \; \partial_t^2 -
                  \partial_x \left(\frac{R^2 u_0^2}{2 u^2} \,\partial_x
\right)
\ee

In the fixed $x$ gauge we have again the same transverse operators, while
the
longitudinal operator becomes
\be
\hat{\cal O}'_x = \frac{5 u_0^6 -2 u^4 u_0^2}{u^6 R^2} +
                  \frac{u_0^2 R^2}{2 u^4} \; \partial_t^2 +
                  \partial_x \left(\frac{u_0^6 R^2}{2 u^8} \,\partial_x
\right)
\ee

Note that we can use the same scaling arguments as for the normal gauge to
show
that the correction is inversely proportional to $L$.
Another remark is in order.
The ``mass term" is more properly looked at as a shift in the levels of
the massless modes rather than as a real mass term,
since it behaves like $u_0^2/R^4$
which is proportional to $1/L^2$.
Thus the mass gap in this case will be of
the same order as the energy of a {\it massless} mode in a box of length
$L$.

\section{  The bosonic string in the \adss background -- Polyakov's action}
\label{sec:polyakov}
The analysis of the quadratic quantum fluctuations can be performed,
as was discussed at the end of section \ref{sec:setup},  also in the
framework of
Polyakov's action.
 Here we present a rederivation of the
bosonic determinant  of the   \adss Wilson loop using the Polyakov action.
Recall that the action \equ{NGAct}   is given
by
\be
        S=\half \int d\sigma d\tau \sqrt{h} h^{\alpha\beta}\partial_\alpha
        X^\mu\partial_\beta X^\nu G_{\mu\nu}(X)
        \label{action}
\ee
with the metric given in \equ{adssmet}.
This action is minimized by the classical configuration
\be
        u(\sigma,\tau)=u_{cl}(\sigma);~~t(\sigma,\tau)=\tau;~~
x(\tau,\sigma)=\sigma;~~
        y^i(\sigma,\tau)=0;~~ \zeta_I(\sigma,\tau)=0
\ee
The function $u_{cl}(\sigma)$ satisfies the equation \cite{Mal2}
\be
        \int_1^{u_{cl}/u_0}{{dy}\over{y^2\sqrt{y^4-1}}} = {{\sigma u_0}\over
R^2}
\ee
where $u_0=u(\sigma=0)$, the minimal value of  $u_{cl}$, is determined by
the boundary
conditions  at $\sigma=\pm L/2$ where $u\rightarrow \infty$, to be
\be
        u_0= \frac{2\sqrt{2}{\pi}^{3/2}R^2}{\Gamma(\frac{1}{4})^2}  \cdot
L^{-1}
\ee
The classical value of the  worldsheet metric is determined by
the  equations of motion derived by variation of the action with respect to
$h_{\alpha\beta}$,
\be
        h_{\alpha\beta}=\partial_\alpha X^\mu \partial_\beta X^\nu
        G_{\mu\nu}(X)
\ee
which take the following form
\be
        ({h_{\sigma\sigma}})_{cl}={u_{cl}^6\over {u_0^4 R^2}}\qquad
        ({h_{\tau\tau}})_{cl}={u_{cl}^2\over R^2}
\ee
The quantum fluctuations of  the string  coordinates are introduced as in
\equ{Xflucts},
 and those of the world sheet metric are as follows
\be
        h_{\sigma\sigma}= (h_{\sigma\sigma})_{cl}(1+\gamma_\sigma);\qquad
h_{\tau\tau}= (h_{\tau\tau})_{cl}(1+\gamma_\tau);\qquad
h_{\sigma\tau}=\gamma
\sqrt{(h_{\sigma\sigma})_{cl}(h_{\tau\tau})_{cl}}
\ee
where  $\gamma, \gamma_\sigma, \gamma_\tau$ parameterize the
metric fluctuations.

We now use the reparameterization invariance of the action to choose the
``gauge''
in which
\be
        \xi^u(\sigma,\tau)=\xi^t(\sigma,\tau)=0
\ee
so that
\be
        u=u_{cl}(\sigma);~~x^t=\tau;~~x^1=\sigma+\xi^1;~~x^2=\xi^2;~~
        x^3=\xi^3
\ee
Note that we do not consider here the fluctuations in the $\zeta$
directions.

Next we  expand the action \equ{action} to quadratic
order in the $\gamma$'s and the $\xi$'s.
The first order correction  of the classical action
$S_{(1)}$ is expected to vanish and indeed
\be
        S_{(1)}={u_0^2\over R^2}\int d\sigma d\tau \partial_\sigma \xi^1
\ee
 vanishes by the boundary conditions on  $\xi^1$.
 The quadratic term is given by:

$$
        S_{(2)}=\half\int d\sigma d\tau \left[
        {u_{cl}^4\over{R^2u_0^2}}\left\{\gamma^2+{1\over 4}\left(\gamma_\tau
        -\gamma_\sigma\right)^2+\left(\partial_\tau\xi^i\right)^2\right\}
        \right.
$$
\be\label{S2action}
        \left.
        +{u_0^2\over R^2}\left\{\left(\partial_\sigma\xi^i\right)^2
        +\partial_\sigma\xi^1\left(\gamma_\tau-\gamma_\sigma\right)\right\}  
      - {{2u_{cl}^2}\over R^2}\gamma\partial_\tau\xi^1
        \right]
\ee
Our  next step is to perform the Gaussian integrals over the $\gamma$'s.
Recall that our goal is to compute
\be
        \czt=\int \cd h_{\alpha\beta}\prod_{i=1}^3 \cd \xi^i
        \exp \left\{ -S_{(2)}\right\}
\ee
where we integrate over the fluctuations of $h_{\alpha\beta}$. This can be
translated
to integrations of the various $\gamma$'s.
The result of the integral is:
\be
        \czt =
        \prod_{i=1}^3 \cd \xi^i
        \exp \left\{ -S_{\rm eff}\right\}
        \label{zteff}
\ee
with
$$
        S_{\rm eff}= \half\int d\sigma d\tau \left[
        {u_0^2\over R^2}\left\{ \left( {u_{cl}^4\over u_0^4}-1\right)\left(
        \partial_\tau\xi^1\right)^2+\left(1-{u_0^4\over u_{cl}^4}\right)
        \left(\partial_\sigma\xi^1\right)^2\right\}
        \right.
$$

\be
       \left.
       +\left\{ {u_{cl}^4\over{R^2u_0^2}}\left(\partial_\tau\xi^\perp\right)^2
       +{u_0^2\over R^2}\left(\partial_\sigma\xi^\perp\right)^2\right\}
       \right]
\ee
where the notation $\xi^\perp$ refers to $\xi^2$ and $\xi^3$.
It is thus clear that in the fixed $u$ gauge
the bosonic operators derived from Polyakov's action are identical to
those found from the NG action \equ{OpAds}.

\section{Introducing fermionic fluctuations}

So far we have considered only  quantum fluctuations of the bosonic
degrees of freedom.
Recall, however, that gauge/gravity duality was originally
\cite{Mal1} proposed in the context of superstring theories which
include worldsheet fermions. This, together with the hope that
supersymmetry will lead to the cancellation of divergences,
raises the question
of the contribution of fermionic fluctuations to the
free energy.  It might seem an easy task to truncate the NSR
action  and collect  terms  quadratic  in the fermionic fluctuations.
This is indeed the case for a  flat background with  vanishing
Ramond--Ramond form.
However there is so far no satisfactory formulation of the
 world-sheet supersymmetric NSR
action for a background with non-vanishing
RR forms. Recall that
in the \adss model, the  self dual 5-form
plays an essential role. Hence we use
here the manifestly space-time supersymmetric GS approach.
For the \adss background this action has been constructed in \cite{MetTse}
as a supersymmetric sigma model with the coset supermanifold
  $SO(2,2|4)/(SO(1,4)\times SO(5))$ as its target space.
The GS action is  invariant under local $\kappa$-symmetry. By gauge fixing
this symmetry one reduces the number of  fermionic degrees of freedom by
a factor of two so that it matches the number of bosonic degrees of freedom.
In \cite{KalTse} a 2d scalar-scalar duality transformation was
invoked to transform the
gauge fixed action  into an action which is quadratic in the fermions.

In this section we first consider the fermionic contribution of the
supersymmetric theory in flat space-time. We then use the
gauge fixed action \cite{KalTse} to deduce the
contribution of the fermionic quantum correction in the \adss background.


\subsection{Fermions in flat space-time}

For flat Minkowski space-time we choose the super-Poincar\'{e}
group as our target space. When we explicitly consider a flat
classical string, the fermionic part of the simplified
gauge fixed GS-action in the flat space-time is simply

\be
    S_F^{flat} = 2 i  \int d\sigma d \tau  \bar \theta  \Gamma^i \pa_i
\theta
\ee
where $\theta$ is a Weyl-Majorana spinor and $\Gamma^i$ are the SO(1,9)
gamma matrices. The fermionic operator is, hence
\be
    \hat{\cal O}_F=D_F = \Gamma^i \pa_i
\ee
and squaring it we get $\Delta = \px^2 - \pat^2$ . The total free energy of
the supersymmetric case is therefore  (using the fact that for D=10, we have
8 transverse coordinates and 8 components of a Weyl-Majorana spinor),

\be
    F = 8 \times \left(- \half \log \det \Delta + \log \det D_F \right) = 0
\ee
and there is no L\"ushcer term.

\subsection{Fermions in \adss}
\label{fermionadss}
The target space for the GS action in the \adss background is
 the coset superspace  $SU(2,2|4) / (SO(1,4) \times
SO(5))$ \cite{MetTse}.
The action is found to be in the form of the
usual supersymmetric string action (but with the induced non--flat metric),
with a change of the derivative acting on the spinor variables
$\theta^1, \theta^2$. If $e_i^a$ is the induced worldsheet vielbien,
and $\omega_i^{a b}$ the induced worldsheet spin connection, then
\be
\label{Di}
\partial_i \theta^I \rightarrow D_i \theta^I \equiv {\cal D}_i \theta^I
- {i \over 2} \epsilon^{I J} e_i^a \gamma^a \theta^J
\ee
with ${\cal D}_i = \partial_i + {1 \over 4} \omega_i^{a b} \gamma_{a b}$
the usual induced worldsheet covariant derivative. The interpretation is
that the metric influences the action through ${\cal D}_i$ while the
Ramond--Ramond flux is responsible for the second term in \equ{Di}.
Using the gauge fixing suggested in \cite{Pesando,KalRah,KalTse}, the
fermionic part of the action for the classical solution can 
be computed and it leads to the  operator
\be
    \hat{\cal O}_F = {u_0^2 \over R^2} \Gamma^1 \px + \left( {u_{cl}^4 \over
{u_0^2 R^2}} \Gamma^0 + {u_{cl}^2 \over R^4} \cdot {\sqrt{u_{cl}^4-u_0^4}
\over u_0^2} \Gamma^2 \right) \pat
\ee
where we use gamma matrices of SO(1,4) which is  the  $AdS_5$ tangent space.
Squaring this operator, we find
\be
    \left( {R^2 \over u_0^2} \hat{\cal O}_F \right)^2 = \px^2 - {u_{cl}^4
\over u_0^4} \pat^2  = {R^2 \over u_0^2} \hat{\cal O}_y
\ee
We find therefore, that the transverse fluctuations partially cancel  the
fermionic fluctuations and we are left with six fermionic degrees of
freedom and the longitudinal and angular bosonic fluctuations.

\section{Bare quark in \adss}
In this section we investigate the single "bare" quark, which is a
flat string in \adss space stretching along the $AdS$ radial coordinate .
In this case the string is a BPS state and
we expect neither its charge nor its mass to be modified by quantum
fluctuations. This problem is related to issues associated with
certain BPS soliton solutions \cite{BPSsoliton}.

The \adss metric is
\be
ds^2 = G_{MN} dx^M dx^N = \left(\frac{R^2}{u^2}\right) du^2 +
                          \left(\frac{u^2}{R^2}\right) dx_\mu dx^\mu +
                          R^2 d\Omega_5^2
\ee
where we take $\{x_0,x_1,x_2,x_3\}$ with Minkowskian signature.

The classical solution of the bare quark is $\sigma=u, \tau=t, x_i=0$
and the $S^5$ angles $\zeta_I$=0. It is obvious that there is a
separation of the quadratic action, and the fluctuations of the
fields can be taken one at a time. We shall take the static
gauge of the action, so that the fluctuating fields are the three
$x_i$ and the five $\zeta_I$.

For the five angular fields, since we are working in the large $R$ regime,
the curvature of $S^5$ is not felt, and we can approximate $S^5$ locally
by the flat space ${\bf R}^5$.
We take one field $\phi(\sigma,\tau)$ along one angle $\zeta$,
i.e, study the worldsheet with
$(u,t,\zeta) = (\sigma, \tau, \phi(\sigma,\tau))$. We have
\be
h_{\alpha \beta} = G_{MN} \partial_\alpha x^M \partial_\beta x^N =
\left(
\begin{array}{cc}
-(u/R)^2 + R^2 \dot{\phi}^2 & R^2 \phi' \dot{\phi} \\
R^2 \phi' \dot{\phi}        & (R/u)^2 + R^2 \phi'^2
\end{array}
\right)
\ee

so that
\be
h = \det h_{\alpha \beta} = -1 + (R^4/u^2) \dot{\phi}^2 - u^2 \phi'^2
\ee
The Nambu--Goto action \equ{NGAct} is therefore
$S \propto \sqrt{-h} = 1 + S_{(2)} + O(\phi^4)$ with the quadratic action
\be
S_{(2)} \propto (R/u)^2 \dot{\phi}^2 - (u/R)^2 \phi'^2 =
\phi \left[ -(R/u)^2 \partial_t^2 + (u/R)^2 \partial_u^2 - 1/R^2 \right]
\phi
\ee
where we have used integration by parts and the Dirichlet boundary
conditions.
We therefore see that for the five angular bosonic fluctuations, the
quadratic operator is
${\cal O} = -(R/u)^2 \partial_t^2 + (u/R)^2 \partial_u^2 - 1/R^2$.

For the three spatial coordinates, we want to keep the fluctuating field
dimensionless, so we take $(u,t,x) = (\sigma, \tau, \phi(\sigma,\tau)/u)$.
A similar calculation yields in this case the operator
${\cal O} = -(R/u)^2 \partial_t^2 + (u/R)^2 \partial_u^2 - 3/R^2$.

Regarding the fermionic fluctuations around the classical bare quark
configuration, one can see that the $\kappa$-symmetry fixing suggested
at \cite{Pesando,KalTse} is not applicable, and produces a degenerate
operator. We should therefore consider only the
quadratic part of the action and try another way to fix the gauge.
The quadratic part of the GS action is, in this case,
\begin {eqnarray}
{\cal L}_F^{(2)} = & & \bar{\theta^0} \left[ - i{R \over u} \gamma^0
  {\cal D}_0 + i {u \over R} \gamma^4 {\cal D}_1 - i {u \over R} \gamma^0
  {\cal D}_1 + i {R \over u} \gamma^4 {\cal D}_0 \right] \theta^0  \non
                 & + & \bar{\theta^1} \left[ - i {R \over u} \gamma^0
  {\cal D}_0 + i {u \over R} \gamma^4 {\cal D}_1 + i {u \over R} \gamma^0
  {\cal D}_1 - i {R \over u} \gamma^4 {\cal D}_0 \right] \theta^1 \\
                 & + & \bar{\theta^0} \left[1 - \gamma^0 \gamma^4
  \right] \theta^1 + \bar{\theta^1} \left[ -1 - \gamma^0 \gamma^4
    \right] \theta^0       \nn
\end {eqnarray}

Fixing the $\kappa$-symmetry by setting $\theta^0 = \theta^1$ and
writing the covariant derivative explicitly,
${\cal D}_0 = \partial_0 + (u/2) \gamma^0 \gamma^4 \;,\;
 {\cal D}_1 = \partial_1$, we have

\be
{\cal O}_F = - {R \over u} \gamma^0 \partial_t + {u \over R}
\gamma^4 \partial_u + {1 \over {2 R}} \gamma^4 + {i \over R} \gamma^0
\gamma^4
\ee

Squaring this operator, integrating by parts and ignoring total
derivatives, we find the operator
\be
{\cal O}_F^2 = -(R/u)^2 \partial_t^2 + (u/R)^2 \partial_u^2 - {{7/4}
  \over {R^2}}
\ee

One should note, however, that taking another gauge fixing, which is
a-priori as good as ours (e.g. $\theta^1 = - i \gamma^3 \theta^0$) leads
to a different result. We prefer our gauge fixing as it leads to the
expected cancelation of the divergences.

All the quadratic operators of the dynamical degrees of freedom are
seen to be of the form
${\cal O}_j = -(R/u)^2 \partial_t^2 + (u/R)^2 \partial_u^2 - c_j/R^2$.
There are eight bosonic ones, corresponding to the eight coordinates
left after fixing the worldsheet reparameterizations, and eight
fermionic ones. The fermionic operators were squared, and are taken
with a negative sign, as the fields are complex Grassmannian.
Therefore, The quadratic quantum contribution to the mass is
\be
\label{Deltam}
\Delta m = -\frac{1}{2 T} \left( 
    \sum_{\mbox{bosons  }} \log \det {\cal O}_j \;\; - 
    \sum_{\mbox{fermions}} \log \det {\cal O}_j   \right)
\ee   

We were unable to find explicitly the eigenvalues of those operators
and multiply them using some regularization scheme. However, it is well
known that the logarithm of the
determinant of such second order
operators has in general quadratic, linear and logarithmic
divergences in the cutoff scale $\Lambda$. Moreover, the corresponding
coefficients are known (see, for example, \cite{FraTse}).
The sheer fact that the number of fermionic and bosonic operators is
the same, and that they differ only in the "mass term" $c_j/R^2$ is
sufficient to ensure that the quadratic and linear divergences cancel
out in \equ{Deltam}.
The logarithmic divergence also cancels out provided that
\be
    \sum_{\mbox{bosons  }} c_j \;\; - 
    \sum_{\mbox{fermions}} c_j  \;\;      =  \;\; 0
\ee   
Indeed, we have $3 \cdot 3 + 5 \cdot 1 - 8 \cdot 7/4 = 0$.
We therefore see that there is no infinite part, needing
renormalization, to the aforementioned contribution to the bare quark
mass. We were unable to show, however, that the finite part also
vanishes, as expected from supersymmetry.

\section{Fluctuations of the $D_p$ backgrounds}
In this section we show that the quadratic contribution of the quantum
fluctuations to the potential is proportional to $1/L$  in the general
Dp background ($p \le 4$). In the D3 case, where the result is dictated by
conformality, this was shown in section \ref{sec:bosonadss}.

For metrics relevant to $Dp$-branes, $-G_{tt} = G_{xx} = G_{yy}$ and
\begin{eqnarray}
f(u) & = & a u^k \\
g(u) & = & b u^j
\end{eqnarray}
with arbitrary $a,b \neq 0 \;,\; k,j$ (for those specific metrics,
$a \sim R^{-(7-p)/2}$). The power-like behaviour of
$f,g$ is essential in order to get the desired result.

The operator for transverse
fluctuations is given by \equ{Oy(u)}. Inserting $f , g$ to that
equation and changing the variable $u$ to $v = u/u_0$ (which has the
range $1 \leq v < \infty$ regardless of $L$), we get
\be
{\cal O}_y =
  \half \left[
      b u_0^j \cdot {{v^j} \over \sqrt{1 - v^{-2k}}} \pat^2 +
      \frac{a^2}{b} u_0^{2k-j-2} \cdot \partial_{v}
        \left( v^{2k-j}\sqrt{1 - v^{-2k}}\partial_{v} \right)
  \right]
\ee
Which is of the form \equ{OAB} with
$A^2 = b u_0^j \;,\; B^2 = \frac{a^2}{b} u_0^{2k-j-2}$. Therefore, by
the result of section \ref{scaling}, the
potential is proportional to $B/A = \frac{a}{b} u_0^{k-j-1}$.
As it is known \cite{KSS2} that
$L \propto \frac{b}{a} u_0^{j+1-k}$, we finally get that $V \propto L^{-1}$,
that is, the
correction to the potential resulting from
transversal bosonic fluctuations (and from all fermionic ones) is
inversely proportional to $L$. Unfortunately, we do not know the
constant of proportionality, and not even its sign.
This constant of proportionality, however, cannot depend on $R$ since
the constants $a,b$ cancel out. Indeed, for $p \neq 3$, the 't Hooft
coupling constant $\lambda \sim R^{7-p}$ is dimensionful, and
therefore can not enter the aforementioned constant of
proportionality. The classical and quantum quadratic potentials add up
to $V \sim - \lambda^{1/(5-p)} / L^{2/(5-p)} + c'/L$ and the quantum
expansion is in $\lambda^{-1/(5-p)}$.

One can easily see that the scaling for the longitudinal bosonic
operator ${\cal O}_x$ in the fixed $u$ gauge is essentially
identical, as it can be casted to the form \equ{OAB} with the same
$A,B$ as in the transverse case.
Therefore we conclude that the correction arising from this gauge fixing 
also obeys $V \propto L^{-1}$.
The picture in the normal gauge, which we argued is safer,
is more involved. We have computed it explicitly and shown the
$L^{-1}$ behaviour only in the $p=3$ case.

\section{Flatness of the string in the confining case}
\label{flat}
\begin{figure}[h!]
\begin{center}
\resizebox{0.8\textwidth}{!}{\includegraphics{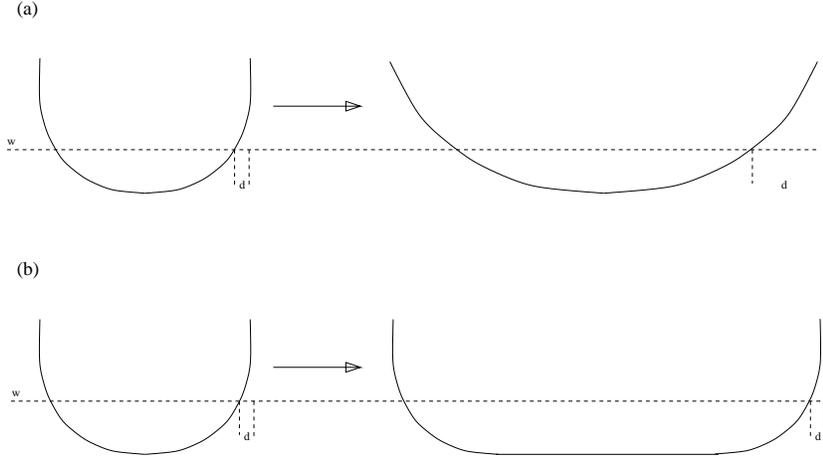}}
\end{center}
\caption{(a) The classical solution is rescaled as $L \rightarrow
  \infty$ in the \adss background while (b) It becomes flatter
  in the confining scenario}
\label{fig:flat2}
\end{figure}

The string in the confining ``pure Yang-Mills'' case was shown
\cite{GrOl2} to be very flat for large $L$. This result will
be needed for the next section.
In this section we shall generalize this result in a precise manner
for all metrics giving confinement.
We shall pick a constant value $w$ (independent of $L$) and inquire
at what distance $d$ from the ends of the string
(at $x = \pm L/2$) the string reaches $u = w$.
We shall find that we cannot show that $d$ is independent of $L$
but rather we find
 that it can grow with $L$. This growth is however mild
and the ratio $d/L$ tends to zero as $L$ grows (see figure \ref{fig:flat2}).

The analysis requires the Taylor expansion of $f(u)$ and $g(u)$
(we use the conventions of \cite{KSS2}),
\bea
\label{ftaylor}
f(u) & = & f(0) + a_k u^k + O(u^{k+1}) \\
\label{gtaylor}
g(u) & = & b_j u^j + O(u^{j+1})
\eea
with $f(0) \neq 0$ due to confinement. We pick $w$ to be sufficiently
small so that all our subsequent approximations are valid.

For small $u$, substitution of \equ{ftaylor},\equ{gtaylor} in \equ{dxucl}
gives
\be
u' \equiv \px u \approx \frac{\sqrt{2 f(0) a_k (u^k - u_0^k)}}{b_j u^j}
\ee
or, with $v \equiv u/u_0$,
\be
v' \approx \frac{\sqrt{2 f(0) a_k}}{b_j} \; u_0^{k/2-j-1} \;
           \frac{\sqrt{v^k - 1}}{v^j}
   \equiv c \; u_0^{k/2-j-1} \; \frac{\sqrt{v^k - 1}}{v^j}
\ee

The critical case $k = 2(j+1)$ is the generic one ($k = 2, \; j = 0$)
and also corresponds to the ``pure Yang--Mills'' case ($k = 1, \; j = -1/2$,
where the horizon $u=u_T$ should be substituted for $u = 0$). In this case
\be
v' \approx c \frac{\sqrt{v^k - 1}}{v^j} \leq c v^{k/2-j} = c v
\ee
so that $(\log v)' = v'/v \leq c$, or $\log v \leq \log v_0 + c x$.
Therefore, $v \leq v_0 e^{c x}$, or finally $u \leq u_0 e^{cx}$.

If we demand $u_0 e^{c x} \leq w$, we get
\be
x \leq \frac{\log (w/u_0)}{c} \equiv \eta - {1 \over c} \log u_0
\ee
On the other hand, it is known \cite{KSS2} that in the critical case
\be
-{1 \over c} \log u_0 = {1 \over 2} L + O(\log L)
\ee
so if $x \le {1 \over 2} L - O(\log L)$ it is guaranteed that $u \le w$.
In other words, $d \le O(\log L)$.

In the non critical case $k > 2 (j+1)$, the minimum of $f(u)$ is more
flat, and we expect the results to be somewhat weaker. Using arguments
similar to those in the previous case (but this time not approximating
$v^k - 1 \approx v^k$), and using results from \cite{KSS2}, one can
show that in this case $d \le O(L^{\frac{k/2-j-1}{k/2-j}})$. $d$ can grow
with $L$, but the ratio $d/L \le O(L^{- \frac{1}{k/2-j}})$ tends indeed
to zero as $L$ grows.

\section{Bosonic  fluctuations in  the "Pure Yang-Mills" setup}
\label{sec:Bosonpym}
When one of the spatial coordinates of the \adss metric is compactified
on a circle of radius $\propto u_T^{-1}$, with the proper boundary
condition \cite{BISY2}, the modes in the compact dimension become
Kaluza-Klein
modes with masses of the order of $u_T$, and the fermions and scalars of
the conformal four-dimensional theory also acquire masses of the same
order and decouple in the low energy regime of the theory.
Therefore, the ${\cal N}=4$ conformal theory in four dimensions
is deformed to a theory similar to non supersymmetric
Yang-Mills in three dimensions.

In the brane language, the metric corresponds to non extremal $D_3$
branes  in  the near horizon limit, and $u_T$ is related to the energy
density on  the $D_3$ brane.

Let us call the compact direction $z$. With the conventions
introduced earlier, working Euclideanly, and still neglecting
factors of $2\pi$, the metric is
\begin{eqnarray}
+G_{tt} = G_{xx} = G_{yy} & = & u^2/R^2 \\
\label{GYM}
                   G_{zz} & = & (1 - (\frac{u_T}{u})^4) \cdot u^2/R^2 \\
                   G_{uu} & = & \left((1 - (\frac{u_T}{u})^4) \cdot
u^2/R^2\right)^{-1}
\end{eqnarray}
and so
\begin{eqnarray}
f(u) & = & u^2/R^2 \\
g(u) & = & (1 - (\frac{u_T}{u})^4)^{-1/2}
\end{eqnarray}

The exact classical solution for the Wilson loop in this setup is
given for this case by \cite{GrOl1,KSS2}
\be
\label{pureYMclassic}
E = \frac{u_T^2}{2\pi R^2} \cdot L -2\kappa + O(e^{-\alpha L})
\ee
with $\kappa$ some constant and $\alpha = 2u_T/R^2$.

In order to compute the quantum correction to the energy, we should
compute the appropriate operators. From \equ{QuadOp} we still have that
\bea
\hat{\cal O}_y & = & \frac{u_0^2}{2R^2} \left[\px^2 + \frac{u^4}{u_0^4}
\pat^2 \right] \\
\hat{\cal O}_\zeta & = & \frac{u_0^2}{2R^2} \left[\px (\frac{R^4}{u^2}
  \px) + \frac{R^4 u^2}{u_0^4} \pat^2 \right]
\eea

but now
\be
\hat{\cal O}_z = \frac{u_0^2}{2R^2} \left[ 
  \partial_x \left( (1 - (\frac{u_T}{u})^4) \partial_x \right) +
  \frac{u^4}{u_0^4} (1 - (\frac{u_T}{u})^4) \partial_t^2 \right]
\ee
A calculation for the longitudinal part using normal fluctuations,
similar to that performed for the conformal case in section \ref{sec:bosonadss},
can
 be carried out, and the result turns out to be similar to the
longitudinal result in that case, only with a different "mass term".

In the large $L$ limit, as we have seen in section \ref{flat}, the
string is, for most of its length, very flat, and has $u \approx u_0$.
Following \cite{GrOl2}, we assume that we can trust this analysis
even after substituting $u \equiv u_0$ in the operators. In that limit
we find
\begin{eqnarray}
\hat{\cal O}_y & \longrightarrow & \frac{u_0^2}{2R^2} \left[ \partial_x^2 +
  \partial_t^2 \right] \\
\hat{\cal O}_\zeta & \longrightarrow & \frac{u_0^2}{2R^2} \left[
  \frac{R^4}{u_0^2} \partial_x^2 + \frac{R^4}{u_0^2} \partial_t^2
\right] \\
\hat{\cal O}_z & \longrightarrow & \frac{u_0^2}{2R^2} (1 -
(\frac{u_T}{u_0})^4)
   \left[ \partial_x^2 + \partial_t^2 \right] \\
\hat{\cal O}_x & \longrightarrow & \left[ \frac{2 u_T^4}{u_0^2 R^4} +
  \frac{1}{2} \partial_x^2 + \frac{1}{2} \partial_t^2 \right]
\end{eqnarray}

As $L$ grows, $u_0 \rightarrow u_T$, and the tendency is exponential
\cite{KSS2},
\be
\label{u0asymp}
\frac{u_0 - u_T}{u_T} \approx e^{-\alpha L}
\ee
Inserting this to the operators we find for large $L$ that
\begin{eqnarray}
\hat{\cal O}_y & \longrightarrow & \frac{u_T^2}{2R^2} \left[ \partial_x^2 +
  \partial_t^2 \right] \\
\hat{\cal O}_\zeta & \longrightarrow & \frac{R^2}{2} \left[ \partial_x^2 +
  \partial_t^2 \right] \\
\hat{\cal O}_z & \longrightarrow & \frac{u_T^2}{2R^2} e^{-2 u_T L}
\left[ \partial_x^2 +
  \partial_t^2 \right] \\
\hat{\cal O}_x & \longrightarrow & \left[ \frac{4 u_T^2}{2R^4} +
  \frac{1}{2} \partial_x^2 + \frac{1}{2} \partial_t^2 \right]
\end{eqnarray}

We see that the operators for transverse fluctuations,
$\hat{\cal O}_y$, $\hat{\cal O}_\zeta$, $\hat{\cal O}_z$,
turn out to be simply the Laplacian in flat spacetime, multiplied by overall
factors, which are, as we have seen, irrelevant. Therefore, the
transverse fluctuations yield the
standard L\"uscher term \cite{LUS} proportional to $1/L$.

The longitudinal normal fluctuations give rise to an operator
$\hat{\cal O}_x$ corresponding to a $1+1$ dimensional scalar field with mass
$2 u_T/R^2 = \alpha$.
Such a field contributes a Yukawa--like term
$\approx -\sqrt{\frac{\alpha}{L}} e^{-2\alpha L}$
to the potential.

We see that the under the assumption that we can
approximate the classical string configuration by a flat one, we get
that one of the bosonic degrees
of freedom becomes massive and its $1/L$ contribution to the potential
is reduced to an exponentially small contribution
(similar to, but even weaker than, the exponential part of the
classical correction  \equ{pureYMclassic}).
This is in contrast to the results of
\cite{GrOl2}, where two bosonic degrees of freedom become massive. The
source of this discrepancy is that in \cite{GrOl2}, the string is
approximated to lie on the horizon, $u \equiv u_T$. Our scheme, in
which the approximation is $u \equiv u_0$, is more accurate.

In the calculations of \cite{GrOl2},
one must take into account the force exerted on the (almost)
flat string by the non flat segments near $x = \pm L/2$. The addition
of the corresponding potential causes the flat string to be in
equilibrium  at $u = u_0$. Therefore, (in the notations of
\cite{GrOl2}), we get $Z^2 + T^2 = u_0 - u_T \neq 0$, and there is a
spontaneous breaking of the rotational symmetry. The resulting Goldstone
boson
is exactly the degree of freedom which is massless in our calculation
but massive in that of \cite{GrOl2}.

\section{ Fermionic fluctuations  in the "pure Yang Mills" setup}
The derivative $D_i$ in \equ{Di} is argued in \cite{MetTse} to
originate from the derivative
\be
\label{Dmu}
D_\mu = \partial_\mu + {1 \over 4} \omega_\mu^{a b} \Gamma_{a b} +
       c \Gamma^{\mu_1 \cdots \mu_5} \Gamma_\mu e^\phi F_{\mu_1 \cdots
\mu_5}
\ee
which appears also in the Killing spinor equation of type IIB supergravity.
 Here, $\phi$ is the dilaton and $F_{\mu_1 \cdots \mu_5}$ is the
Ramond--Ramond field strength.
In the \adss case, there is no dilaton background
($\phi = 0$), and
$F_{\mu_1 \cdots \mu_5} \sim u^3 \epsilon_{\mu_1 \cdots \mu_5}$.
In the actual computation, a factor of $(-\det G_{\mu \nu})^{-1/2} \sim
u^{-3}$
cancels the $u^3$ from $F$, and
the last term of \equ{Dmu} indeed reduces to the last term of \equ{Di},
with the Minkowski--space $\gamma^{01234}$ replaced by $\epsilon^{I J}$.

When moving to the "pure Yang--Mills" metric \equ{GYM},
upon compactification of the coordinate $z$, the dilaton and
Ramond--Ramond flux remain the same, and $\det G_{\mu \nu}$
does not change either. Assuming the same form
\equ{Dmu} of the derivative, we therefore again arrive at \equ{Di}.

Based on the results of section \ref{flat}, We now assume, as in the
bosonic computation of the previous section, that the classical string
is flat. That is,
$\sigma^0 = t, \sigma^1 = x^1$, with $u = u_0$ and all other
coordinates zero (or constant). The combined
fluctuations of the string around its classical solution and of the
fermions around zero give terms of high order; the quadratic terms of the
fermions are coupled only to the bosonic classical solution.

In this section, we find it more
convenient to work in Euclidean coordinates.
It is straightforward to see that the induced vielbein $e_i^a$ obeys
$e_0^0 = e_1^1 = u_0/R$, and all the other components vanish. The
induced metric is therefore proportional to the identity matrix, as it
clearly should, and the Lagrangian reduces to
\newcommand{\LTT}[4]{\mbox{$  \bar{\theta}^#1 \gamma^#2 D_#3 \theta^#4 $}}
\begin{eqnarray}
\label{Lflatferm}
{\cal L} & \sim & \mbox{} + \LTT{1}{0}{0}{1} + \LTT{2}{0}{0}{2} +
\LTT{1}{1}{1}{1} + \LTT{2}{1}{1}{2} \\
         &      & \mbox{} - \LTT{1}{0}{1}{1} + \LTT{2}{0}{1}{2} +
\LTT{1}{1}{0}{1} - \LTT{2}{1}{0}{2} \nn
\end{eqnarray}
with $D_i$ the covariant derivative containing the effects of gravity
and of the Ramond--Ramond flux.

The relevant components of the spin--connection induced on the
classical worldsheet are, by \equ{u0asymp},
\be
\label{wflatferm}
2 \epsilon \equiv \omega_0^{0 4} = \omega_1^{1 4} =
\frac{u_0}{R^2} \sqrt{1 - (u_T/u_0)^4} \sim \frac{2 u_T}{R^2} e^{-(u_T/R^2) L}
\ee
By the previous discussion, the term arising from the Ramond--Ramond
flux in $D_i$ is $c \gamma^i \epsilon^{I J}$. In order to agree with
\equ{Di}, $c = u_0/2R^2$. In conclusion,
\bea
\label{D0}
D_0 \theta^I & = & \left[ (\partial_0 + \epsilon \gamma^0 \gamma^4)
                          \delta^{I J}  +
                          c \gamma^0 \epsilon^{I J}                 \right]
                   \theta^J \\
\label{D1}
D_1 \theta^I & = & \left[ (\partial_1 + \epsilon \gamma^1 \gamma^4)
                          \delta^{I J}  +
                          c \gamma^1 \epsilon^{I J}                 \right]
                   \theta^J
\eea

Inserting \equ{D0},\equ{D1} into \equ{Lflatferm} we get $24$ terms
which reduce to
\bea
{\cal L} \;\; \sim &   & \bar{\theta}^1 \left[ 
                   \gamma^0 \partial_0 + \gamma^1 \partial_1 
                 - \gamma^0 \partial_1 + \gamma^1 \partial_0 
                 + 2 \epsilon \gamma^4 - 2 \epsilon \gamma^0 \gamma^1 \gamma^4
                            \right] \theta^1 \non
                   & + & \bar{\theta}^2 \left[ 
                   \gamma^0 \partial_0 + \gamma^1 \partial_1 
                 + \gamma^0 \partial_1 - \gamma^1 \partial_0 
                 + 2 \epsilon \gamma^4 + 2 \epsilon \gamma^0 \gamma^1 \gamma^4
                            \right] \theta^2 \\
                   & + & \bar{\theta}^1 \left[ +2c - 2c \gamma^0 \gamma^1 \right] 
                    \theta^2
                 \; + \; \bar{\theta}^2 \left[ -2c - 2c \gamma^0 \gamma^1 \right]
                    \theta^1 \nn
\eea

Now we should impose a $\kappa$--symmetry gauge. Choosing the gauge 
$\theta^2 = -i \gamma^4 \theta^1$, as advocated by \cite{KalTse} for the
\adss case, 
we get
\be
{\cal L} \sim \bar{\theta}^1 \left[ \gamma^0 \partial_0 + \gamma^1 \partial_1                             + 2i (c-i\epsilon) \gamma^0 \gamma^1 \gamma^4
                             \right] \theta^1
         \equiv \bar{\theta}^1 {\cal O}_F \theta^1
\ee
When we square the fermionic operator, and remove, using integration by
parts, the terms with a single derivative, we finally get
\be
{\cal O}_F^2 = \partial_0^2 + \partial_1^2 + 4 (c - i \epsilon)^2
\ee
We thus see that all the eight fermionic modes become massive, with
mass $m = 2 (c - i \epsilon)$. The mass of the fermions has an imaginary
part,
but as by \equ{wflatferm},
$\epsilon$ is vanishingly small for large distances $L$, we see that
$m \approx 2c = u_T/R^2$ is half the mass of the only massive bosonic
mode.

Choosing the similar $\kappa$--symmetry gauge
$\theta^2 = -i \gamma^3 \theta^1$
we get a similar result,
\be
{\cal L} \sim \bar{\theta}^1 \left[ \gamma^0 \partial_0 + \gamma^1
\partial_1
                                    + 2 \epsilon \gamma^4
                                    + 2i c \gamma^0 \gamma^1 \gamma^3
                             \right] \theta^1
\ee
Now
\be
{\cal O}_F^2 = \partial_0^2 + \partial_1^2 + 4 (c^2 + \epsilon^2)
\ee
and we have
$m^2 = 4 (c^2 + \epsilon^2) \approx 4 c^2$,
so again $m \approx 2c$ in the large $L$ limit.

In conclusion, we see that in the "pure YM" case there are seven
massless bosonic modes, and no
fermionic ones. The contribution of the L\"{u}scher term to the quark
anti--quark potential is $-(7-0) \cdot \pi/24 \cdot 1/L$,
which is attractive and concave in agreement with the general
result in quantum field theory \cite{Bachas,DorPer}.

\section{Summary and discussion}

Wilson loops are some of the most important gauge invariant physical
quantities associated with non-Abelian gauge theories. They play an
essential role in the understanding of the underlying structure of YM
theory. In particular, the rectangular loop has a simple
interpretation, being related to
the static potential between a quark and an
anti-quark. Wilson loops have thus attracted a tremendous amount of work
throughout the years. Among the various approaches invoked to investigate
the Wilson loop, the description in terms of a string model is the
most promising. This approach is based on the fact that stringy Wilson
loops obey the loop equation, and on the phenomenological picture of a
flux tube connecting the quark pair. Recently the duality conjecture of
Maldacena \cite{Mal1} has given this approach new impetus from a
new perspective.

It is easy to realize that a classical string in flat space--time
 results in an area law behaviour  which is expected for a
 gauge theory in the  confining phase. An interesting issue that had
 been discussed in the early Eighties is the quantum
 corrections to the linear potential. The purpose of the present work
 is to shed additional light on this issue, using the gravity
 description of large N gauge theories. En route to this goal we have
also addressed the issues of the quantum correction to the quark
anti--quark potential in ${\cal N}=4$ SYM via quadratic fluctuations
 of the corresponding string in the \adss background, as well as the
 contribution of these fluctuations to the mass of the BPS quark in
 that approach.

Determining the quantum corrections involves four steps:  (i) Writing
down the corresponding action, (ii) Choosing an appropriate gauge
fixing, (iii) Expanding the gauge fixed action to quadratic order and
(iv) Computing, using some regularization scheme, the functional
determinant and the resulting free energy. Whereas the bosonic part of
the action is well known for any background, the incorporation of
fermionic degrees of freedom is more involved, due to the presence of
a non trivial Ramond--Ramond background in the models we discuss. So far no
complete NSR action for such systems has been formulated. A
Green--Schwarz type of action was formulated for the \adss model
\cite{MetTse}, but not for the more general backgrounds that
correspond to confining scenarios.
In attempt to generalize the fermionic action we used the fact that the
\adss action is a covariantization of the flat GS action, and,
assuming the same behaviour, wrote down the fermionic operators for
the confining scenario.

It turned out that the issue of choosing a "gauge slice" is more tricky.
Already in fixing the 2d reparameterization we found that different
gauge choices lead to different operators. In principle, different
operators could yield the same regularized determinant. Moreover,
since we were not able to compute explicitly the determinants, we
cannot really claim that they are indeed different. However even
the argument that the fixed $u$ gauge, the fixed $x$ gauge
and the "normal gauge" lead to the same free energy was
plagued with possible singularities. We also argued that
the latter gauge is safer than the others. The projection of the
fermionic fields into 8 Grassmannian degrees of freedom (a single
Weyl-Majorana spinor) through fixing of the $\kappa$--symmetry can also be
performed in various ways, and again, different gauge slices yield
different operators. For the BPS single quark configuration of the
\adssns, the $\gamma_4$ gauge fixing, introduced in \cite{KalTse}, was
found to be inapplicable (as in general, fixing the $\kappa$-symmetry
using a $\gamma_i$ projection cannot be used when the classical
configuration spans along $x_i$) and of the other projections we tried
one lead to a cancelation of the logarithmic divergences while another
did not. Further investigation is required in order to gain better
understanding of the issue of choosing a legitimate and effective
$\kappa$-symmetry gauge fixing.

As for  step (iv), we were not able to compute explicitly the
determinants, apart from the known case of flat space-time. Regarding
the confining setup, we used similar considerations to those of
\cite{GrOl2} in treating the bosonic degrees of freedom and found
that (only) one mode becomes massive. We were also able to find some
evidence that the fermionic modes are massive as well, thus leading to
a quantum correction to the linear potential which is attractive and
of L\"uscher type. This provides an evidence in favour of the
gauge/gravity duality, as it was shown in \cite{Bachas}, based on
general arguments, that the overall quantum potential should be concave.

The ultimate test of the results concerning the confining setups is
comparison with experimental observations. There were several attempts
to deduce the corrections to the linear potential from the spectrum of
heavy mesons but it is not clear to us whether one can reach a coherent
picture from this analysis. On the way to comparing with experiment
one should also consider lattice computations of Wilson loops in 3 and
4 dimensional pure YM theory. In  particular, one should address the
question whether a L\"uscher term correction to the linear potential
can be seen in lattice simulations. According to \cite{Teper}
there are some numerical evidence for a L\"uscher term associated with a
bosonic string. However the results are not precise enough to be
convincing.

It is the  quantum fluctuations of the \adss case that leave us with
several unresolved puzzles. In \cite{FGT} it was found that there
is a logarithmic divergence in the determinant and it was suggested
that the infinity should be canceled by renormalizing the mass of the
 bare quarks. However, the latter are BPS states and thus should not
 receive any quantum corrections. Indeed, we have shown that, at least
 in a certain $\kappa$-symmetry gauge fixing there are no
logarithmic divergences. On the other hand we could not find any gauge
fixing in which  the  quantum corrections to the
Wilson loop are free of
logarithmic divergences. Hence, the renormalization and gauge fixing
procedures for this case deserve further investigation.
As  for the  question of whether the \adss case admits a L\"uscher term
 and of what sign, we again do not have a conclusive answer.
There are no hints from the explicit calculation that the bosonic
and fermionic determinants cancel. As discussed in the introduction,
the coefficient of the L\"uscher term is in fact related to the "intercept",
the normal ordering constant in the Virasoro $L_0$ generator.
We do not have any evidence that this should vanish in the $AdS$ case.

There is, however, a naive argument why the L\"uscher term should not exist.
In the limit of $R\rightarrow \infty$ one may speculate that locally one
sees a flat metric and hence there should be a vanishing coefficient.
Now, since the basic property of the L\"uscher term is that it is
independent
on $R$, the result for $R\rightarrow \infty$ should hold for any
$R$. As we couldn't verify this prediction we may suspect that the
argument is too naive and one cannot extrapolate smoothly to the
Wilson loop in flat space time.
It is  thus clear that the explicit evaluation of the various determinants
in the \adss case is still an open and challenging question.

\vspace{12pt}
{\bf Acknowledgements}

We have greatly benefitted from discussions with O. Aharony,
R. L. Jaffe, S. Theisen, A. A. Tseytlin and S. Yankielowicz.

\end{document}